\documentclass[10pt,aps,physrev,raggedbottom,longbibliography,nobalancelastpage,reprint,citeautoscript,letterpaper,superscriptaddress,onecolumn]{revtex4-2}
\usepackage{braket}
\usepackage{graphicx} % Required for inserting images
\usepackage{subcaption} 
\usepackage{amsmath}    % For mathematical equations
\usepackage{amssymb}    % For mathematical symbols
\usepackage{hyperref}   % For clickable references
\usepackage{float}      % For figure positioning
\usepackage{siunitx}
\begin{document}

\title{Accelerated characterization of two-level systems in superconducting qubits via machine learning}

\author{Avinash Pathapati}
\affiliation{Nordita,
Stockholm University and KTH Royal Institute of Technology,
Hannes Alfvéns väg 12, SE-106 91 Stockholm, Sweden}

\affiliation{Department of Physics, University of Connecticut, Storrs, Connecticut 06269, USA}

\author{Olli Mansikkamäki}
\affiliation{Nordita,
Stockholm University and KTH Royal Institute of Technology,
Hannes Alfvéns väg 12, SE-106 91 Stockholm, Sweden}

\author{Alexander Tyner}
\affiliation{Nordita,
Stockholm University and KTH Royal Institute of Technology,
Hannes Alfvéns väg 12, SE-106 91 Stockholm, Sweden}
\affiliation{Department of Physics, University of Connecticut, Storrs, Connecticut 06269, USA}

\author{Alexander Balatsky}
\affiliation{Nordita, 
Stockholm University and KTH Royal Institute of Technology,
Hannes Alfvéns väg 12, SE-106 91 Stockholm, Sweden}
\affiliation{Department of Physics, University of Connecticut, Storrs, Connecticut 06269, USA}

\begin{abstract}
We introduce a data-driven approach for extracting two-level system (TLS) parameters—frequency \(\omega_\text{TLS}\), coupling strength \(g\), dissipation time  \(T_{\text{TLS}, 1}\), and the pure dephasing time \(T^\varphi_{\text{TLS}, 2}\), labelled as a 4 component  vector $\vec{q}$, directly from simulated spectroscopy data generated for a single TLS by a form of two-tone spectroscopy. Specifically, we demonstrate that a custom convolutional neural network model(CNN) can simultaneously predict \(\omega_\text{TLS}\), \(g\), \(T_{\text{TLS}, 1}\) and \(T^\varphi_{\text{TLS}, 2}\) from the spectroscopy data presented in the form of images. Our results show that the model achieves superior performance to perturbation theory methods in successfully extracting the TLS parameters. Although the model, initially trained on noise-free data, exhibits a decline in accuracy when evaluated on noisy images, retraining it on a noisy dataset leads to a substantial performance improvement, achieving results comparable to those obtained under noise-free conditions. Furthermore, the model exhibits higher predictive accuracy for parameters  \(\omega_\text{TLS}\) and \(g\) in comparison to  \(T_{\text{TLS}, 1}\) and \(T^\varphi_{\text{TLS}, 2}\).

\end{abstract}

\maketitle

\section{Introduction}

Superconducting qubits offer one of the most promising pathways to build scalable quantum computers. To further advance the performance of qubit based devices, one would need to understand and mitigate the effects that limit the performance. Major limitations for the performance of qubits arise from short coherence times, and this effect is amplified in a quantum chip with large number of qubits. One of the main contributors affecting coherence times has been attributed to the two-level system (TLS) present in the building blocks of the qubits~\cite{kjaergaard_superconducting_2020, wang_surface_2015, gambetta_investigating_2017, barends_coherent_2013, muller_towards_2019, klimov_fluctuations_2018, burnett_decoherence_2019, schlor_correlating_2019, spiecker_two-level_2023, thorbeck_readout-induced_2024,tyneridentification}. Real-time identification of qubit parameters affected by TLSs, particularly in quantum processors with a large number of qubits, is exceedingly difficult. Streamlining the process of identifying these decoherent qubits will help in producing better quantum chips. 

Deep learning \cite{lecun2015deep} has revolutionized various fields, with convolutional neural networks (CNNs) being particularly successful in tasks related to computer vision. CNNs can automatically extract relevant features from images through a sequence of convolution and pooling operations when trained against a required objective. In our case, the input is the spectroscopy data of simulated qubits represented as images and the objective is to predict the four TLS parameters denoted by $\vec{q} = [ \omega_\text{TLS}, g, T_{\text{TLS}, 1},\, T^\varphi_{\text{TLS}, 2}]$. Here, $\omega_\text{TLS}$ is the TLS frequency, $g$ is the coupling strength, $T_{\text{TLS}, 1}$ is the dissipation time, and $T^\varphi_{\text{TLS}, 2}$ is the pure dephasing time.
More information on the working of CNNs can be found in e.g., \cite{lecun1995convolutional}. In our earlier work \cite{mansikkamaki2024two}, we developed a two-tone spectroscopy protocol for frequency-fixed qubits and proposed a specific method to extract $\omega_\text{TLS}$ from two-dimensional intensity maps obtained by varying both the drive frequency and pulse duration. As a proof-of-principle, a simple CNN model was used as a benchmark for estimating $\omega_\text{TLS}$. For more information on this previous work, please see \ref{Methods}.
 
 In this work, we extend this approach by proposing a deep learning model that can extract all the TLS parameters $\vec{q}$ for a single TLS.
 
The model we put forth in this work is based on ResNet\cite{he2016deep}. ResNet is an advanced CNN architecture that was shown to successfully alleviate the problem of vanishing gradients in deep networks \cite{he2016deep} during training. ResNet introduced the concept of residual learning, where identity mappings (skip connections) allow gradients to flow more effectively during backpropagation. This trick enables the training of extremely deep networks, significantly improving the performance on image classification benchmarks such as ImageNet \cite{deng2009imagenet}. In this paper, we built a custom architecture based on ResNet that can extract $\vec{q}$ from the spectroscopy images. This method enables swift, automated characterization of faulty qubits and is particularly effective in identifying faulty ones among large qubit arrays, thereby accelerating the cycle of designing better quantum chips. 

\section{Method} \label{Methods}

Deep learning models require immense datasets for training. The dataset used for training the model at hand consisted of two-dimensional $(\omega, t)$ spectroscopy maps from our previous work \cite{mansikkamaki2024two}, where $\omega$ is the drive frequency and $t$ is the pulse duration. These maps were generated by numerically solving the Lindblad master equation for different qubit and TLS parameters, with random Gaussian noise added to emulate experimental conditions. The complete dataset construction process is detailed in \ref{dataset}.For the model architecture, ResNet was used as a feature extractor, followed by several fully connected layers (FC) at the end to predict $\vec{q}$. Further information on the model architecture and the training procedure is provided in \ref{"model_arch"} and \ref{training_proc} respectively.

\subsection{Dataset}\label{dataset}
\par 
As stated, the model is trained to analyze data generated via two-tone spectroscopy protocol introduced in Ref. \cite{mansikkamaki2024two}. This protocol was put forth to directly determine TLS frequencies and the strength of coupling to the qubit. It is explained in detail in Ref. \cite{mansikkamaki2024two}, but is summarized here for clarity. 
\par 
First, we prepare the system in its ground state, $\ket{00}$, by waiting until the population is fully dissipated. The notation $\ket{ij}$ refers to $i = 0,1,2$ for qubit state and second index refers to $j = 0,1$ state of TLS. Second, the transmon is driven with the drive frequency $\omega_d$ until time $t_A$. We label this pulse A. Third, a $\pi$-pulse of length $t_\pi$ is performed at the measured frequency $\tilde \omega_q$. We label this pulse B. Last, the transmon population is measured. Due to the changing TLS population, the post-pulse B population of the transmon at time $t_B = t_A + t_\pi$ will vary depending on the time $t_A$.
\par
\begin{figure}[h]
    \centering
    % Subfigure 1
    \begin{subfigure}{0.45\textwidth}  % Adjust width as needed
        \centering
        \includegraphics[width=\textwidth]{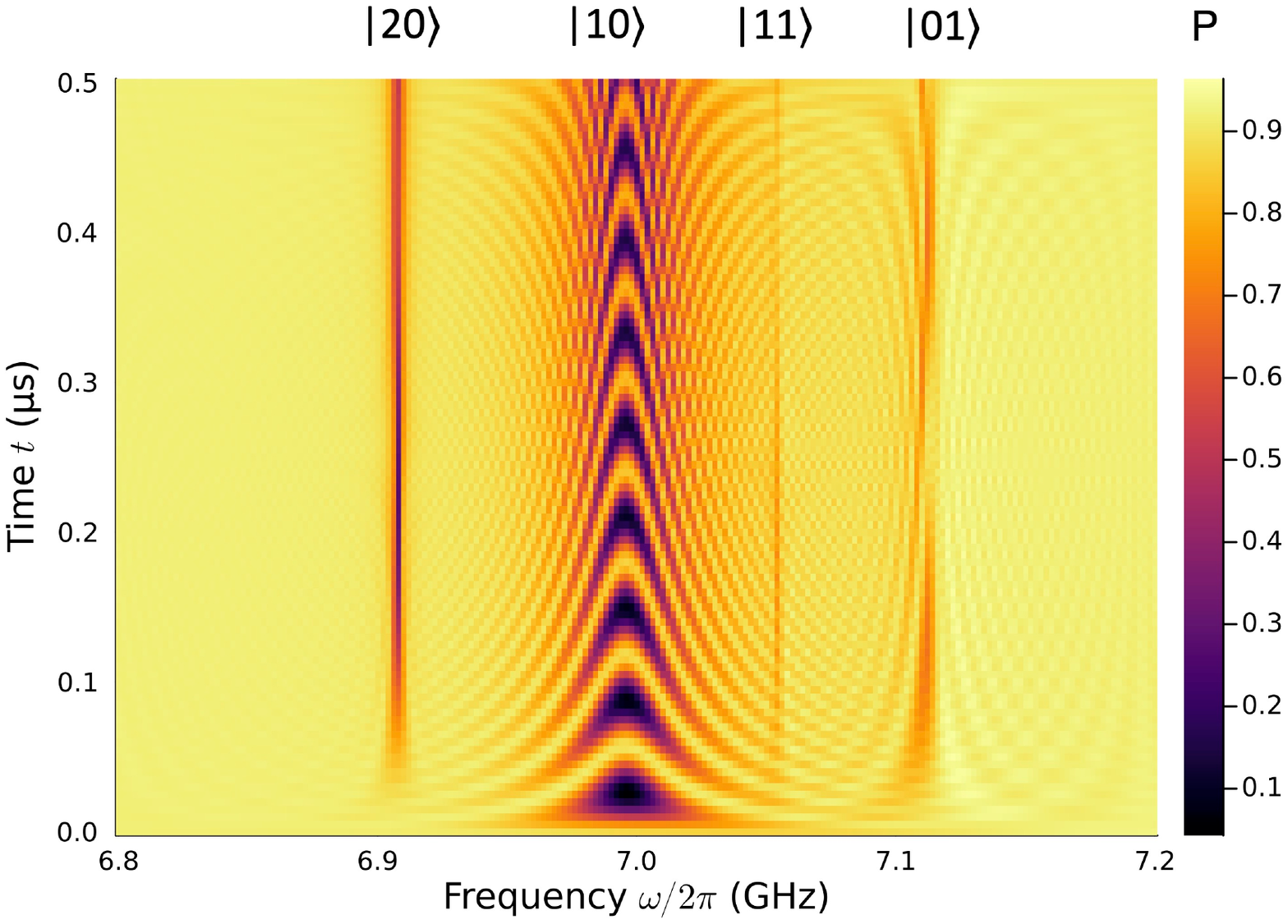}  
        \label{fig:no_noise_image}
        \label{fig:}
    \end{subfigure}
    \hfill  
    \begin{subfigure}{0.45\textwidth}
        \centering
        \includegraphics[width=\textwidth]{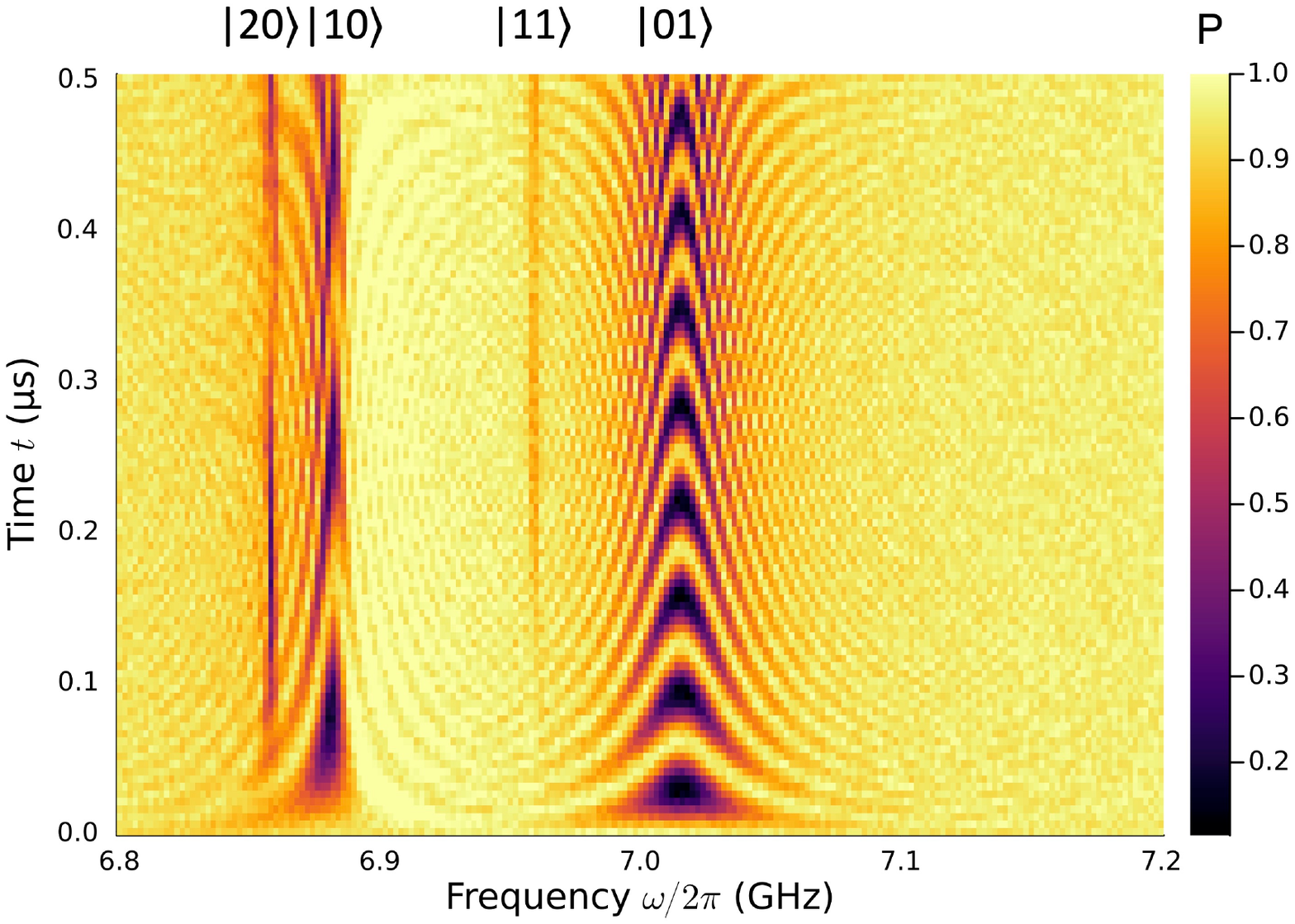} 
        \label{fig:noise_image}
    \end{subfigure}
    \caption{\textbf{Synthetic spectroscopic training data.} Examples of synthetic spectroscopy data generated via the Lindblad master equation, eq. \eqref{eq:master}. The data is represented as an image such that distinct states, labeled $|nm\rangle$, are identifiable. This data is generated considering (a) no noise and (b) noise, modeled as a random shift to the qubit population.}
    \label{fig:tls_images}
\end{figure}

 
These steps are repeated for a range of drive frequencies $\omega_d$ and lengths $t_A$ of the pulse A. The population $P$ of the first excited level of the transmon after pulse B as a function of the pulse A frequency $\omega_d$ and length $t_A$ are then mapped to create the images shown in Fig. \eqref{fig:tls_images}. The labels mark the state $\ket{nm}$ of the $\ket{00} \leftrightarrow \ket{nm}$ transition represented by the feature. The $ \ket{10}$ and $ \ket{20}$ states correspond to the first two excited states of the transmon. The TLS will have three additional transitions: from the ground state to $\ket{01}$, $\ket{11}$, and $\ket{21}$. The last of these is generally slow compared to typical dissipation times and ignored here. The $\ket{01}$ corresponds to the two-qubit cross-resonance gate~\cite{rigetti_fully_2010, chow_simple_2011}. The $\ket{11}$ transition corresponds to the bSWAP two-qubit gate~\cite{poletto_entanglement_2012}. 
\par
The parameters of the TLSs, i.e., the frequencies $\tilde \omega_k$, the coupling strengths $g$, and the dissipation and pure dephasing times $T_{1, k} = 1 / \gamma_k$ and $T_{\varphi, k} = 2 / \kappa_k$, all impact the resulting data shown in Fig. \eqref{fig:tls_images}, creating the opportunity for the machine learning network to extract the TLS parameters and the motivation to use these images as training data. 

\par

The training data was generated with the Lindblad master equation
\begin{align}\label{eq:master}
    \begin{split}
        \frac{d \hat \rho}{d t} =& -\frac{i}{\hbar} [\hat H, \hat \rho] \\
        &-\frac{\gamma_q}{2} (\hat a^\dagger \hat a \hat \rho + \hat \rho \hat a^\dagger \hat a - 2 \hat a \hat \rho \hat a^\dagger) \\
        &-\frac{\kappa_q}{2} (\hat a^\dagger \hat a \hat a^\dagger \hat a \hat \rho + \hat \rho \hat a^\dagger \hat a \hat a^\dagger \hat a - 2 \hat a^\dagger \hat a \hat \rho \hat a^\dagger \hat a) \\
        &-\frac{\gamma_\text{TLS}}{2} (\hat b^\dagger \hat b \hat \rho + \hat \rho \hat b^\dagger \hat b - 2 \hat b \hat \rho \hat b^\dagger) \\
        &-\frac{\kappa_\text{TLS}}{2} (\hat b^\dagger \hat b \hat b^\dagger \hat b \hat \rho + \hat \rho \hat b^\dagger \hat b \hat b^\dagger \hat b - 2 \hat b^\dagger \hat b \hat \rho \hat b^\dagger \hat b),
    \end{split}
\end{align}
where $\hat \rho$ is the density matrix describing the state of the system, $\hat a$ and $\hat b$ are the bosonic annihilation operators for the qubit and the TLS, respectively. $\gamma_{q/\text{TLS}} = 1 / T_{q/\text{TLS}, 1}$ and $\kappa_{q/\text{TLS}} = 1 / T^\varphi_{q/\text{TLS}, 2}$ are the dissipation and pure dephasing rates. The Hamiltonian describing the coherent time evolution is given by
\begin{equation}
    \hat H = \omega_q \hat a^\dagger \hat a - \frac{U}{2} \hat a^\dagger \hat a^\dagger \hat a \hat a + \omega_\text{TLS} \hat b^\dagger \hat b + g (\hat a^\dagger \hat b + \hat b^\dagger \hat a),
\end{equation}
where $\omega_q$ is the qubit frequency, $\omega_\text{TLS}$ is the TLS frequency, $U$ is the anharmonicity of the qubit, and $g$ is the coupling strength. We limit the maximum population of the qubit to two and, as the name suggests, the population of the TLS to one. 

Since the detuning $\Delta = \omega_\text{TLS} - \omega_q$ is what matters for the physics of the system, we set $\omega_q / 2 \pi = \SI{7}{\giga \hertz}$. All other parameters are randomized and picked from uniform distributions: $\omega_\text{TLS} \in [\SI{6.75}{\giga \hertz}, \SI{7.25}{\giga \hertz}]$, $U / 2 \pi \in [\SI{150}{\mega \hertz}, \SI{300}{\mega \hertz}]$, $g / 2 \pi \in [\SI{5}{\mega \hertz}, \SI{50}{\mega \hertz}]$, $T_{q, 1} \in [\SI{1}{\micro \second}, \SI{10}{\micro \second}]$, $T^\varphi_{q, 2} \in [\SI{0.5}{\micro \second}, \SI{5}{\micro \second}]$,  $T_{\text{TLS}, 1} \in [\SI{0.5}{\micro \second}, \SI{10}{\micro \second}]$, and $T^\varphi_{\text{TLS}, 2} \in [\SI{0.5}{\micro \second}, \SI{20}{\micro \second}]$. 

The above Lindblad equation models the average time-evolution of a quantum mechanical system under dissipation and dephasing, and therefore predicts a continuous evolution of the qubit population $P = \langle \hat a^\dagger \hat a \rangle = \text{Tr}(\hat \rho \hat a^\dagger \hat a)$. In reality, unless the population of the qubit is averaged over a large number of measurements, one would expect to see sharp jumps in the population at essentially random times, resulting in a noisy evolution. Additionally, an experimentally realistic setup would have additional sources of noise, e.g., in the power and frequency of the drive and measurement errors, that are not captured by our model. Noting that CNNs trained on non-noisy data perform badly on noisy data, we have included a simple noise term on the computed population $\tilde P = P + \delta P$, where $\delta P$ is randomly picked from an uniform distribution. The width of the distribution is varied from $0.01$ to $0.2$ for each sample to represent experimental setups of varying quality. 

Figure \ref{fig:tls_images} provides representative examples of spectroscopy images, with and without noise from the dataset. The data discussed in this work always contains only a single strongly coupled TLS to accommodate rapid generation of a large dataset. We briefly outline the steps needed to generate the $(\omega, t)$ maps from Figure \ref{fig:tls_images} as follows: prepare the system in its ground state $|00\rangle$, e.g., by waiting until the population is fully dissipated, then drive the transmon with the drive frequency $\omega_d$ until time $t_A$. We label this pulse A, perform a $\pi$-pulse of length $t_\pi$ at the measured frequency $\tilde{\omega}_q$. We label this pulse B and finally measure the transmon population. Due to the changing TLS population, the post-pulse B population of the transmon at time $t_B=t_A+t_\pi$ will vary depending on the time $t_A$;

\subsection{CNN model}\label{"model_arch"}

\begin{figure}[h] 
    \centering
    \includegraphics[width=0.6\textwidth]{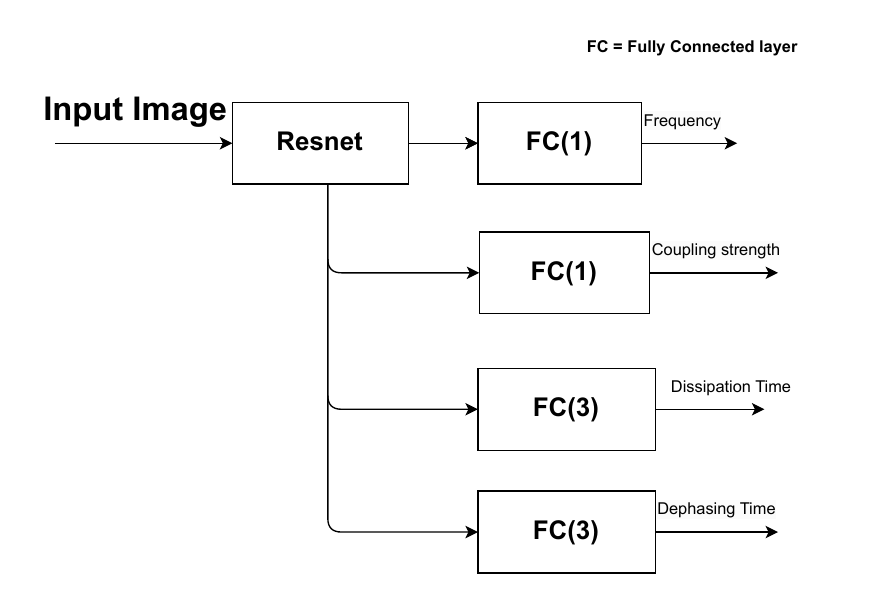}  

    \caption{\textbf{Architecture of CNN model.} Schematic architecture of the  model used for predicting $\vec{q}$}
    \label{fig:tls_dl_model}  
\end{figure}

The architecture of the custom CNN model is illustrated in Figure~\ref{fig:tls_dl_model}. As discussed, the CNN model employs ResNet as the first layer, extracting the required features from the images. This initial layer is followed by four, independent, fully connected (FC) blocks. Each block consists of multiple FC layers arranged sequentially, with the final output from the block corresponding to one component of $\vec{q}$. The FC layers used inside each block transform the output from the ResNet layer into $\vec(q)$ through a series of linear and non-linear operations. Each FC layer performs this step \[
\mathbf{y} = \sigma(\mathbf{W} \mathbf{x} + \mathbf{b}),
\]

where \( \mathbf{x} \) is the input vector, \( \mathbf{W} \) and \( \mathbf{b} \) are the learned weights and biases, respectively, and \( \sigma \) denotes a non-linear activation function. The final output of the network is the parameter vector \( \vec{q} \). The notation \texttt{FC(\emph{n})} in the figure \ref{fig:tls_dl_model} represents a stack of \emph{n} fully connected layers, applied sequentially. This compact notation is used to indicate depth within each output branch of the network.

\subsection{Training procedure}\label{training_proc}
The dataset is partitioned into three subsets: a training set (35,000 images), a validation set (15,000 images), and a test set (5,000 images). The output vector $\vec{q}$ was normalized column-wise to the range $[1, 10]$ using the transformation
\begin{equation}
q^{(i)}_{\mathrm{norm}} = 1 + 9 \cdot \frac{q^{(i)} - q^{(i)}_{\min}}{q^{(i)}_{\max} - q^{(i)}_{\min}},
\end{equation}
where $q^{(i)}$ denotes the values in the $i$-th column of the vector $\bar{q}$, and $q^{(i)}_{\min}$ and $q^{(i)}_{\max}$ are the corresponding minimum and maximum values within that column. 
This scaling ensures scale-independent learning by preventing parameters with larger magnitudes from disproportionately influencing the loss. Model training is conducted using the training set, while performance on the validation set is monitored to guide the training process. Specifically, we employ an \textit{early stopping} criterion, whereby training is terminated when no further improvement is observed in the validation loss. This approach helps to prevent overfitting and ensures better generalization to unseen data.

The model is trained to minimize the mean squared error (MSE) between the predicted and actual values of the output vector \( \vec{q} = [\omega_{\text{TLS}}, g, T_{\text{TLS}, 1}, T^{\phi}_{\text{TLS},2}] \)

The loss function is defined as:
\[
\mathcal{L} = \frac{1}{4*N} \sum_{i=1}^{N} \left\| \vec{q}_i - \hat{\vec{q}}_i \right\|^2
\]
where \( N \) is the number of training samples, \( \vec{q}_i \) is the ground truth vector for the \( i \)-th sample, and \( \hat{\vec{q}}_i \) is the corresponding prediction by the model.

\section{Results}
 
Model \textbf{M1} is trained under noise-free conditions i.e. $\delta P = 0$, Model \textbf{M2} is trained on a noise-free dataset ($\delta P = 0$) but tested on the noisy dataset and Model \textbf{M3} is trained and tested on the noisy dataset.

\begin{figure}[h]
    \centering
    \includegraphics[width=0.8\linewidth]{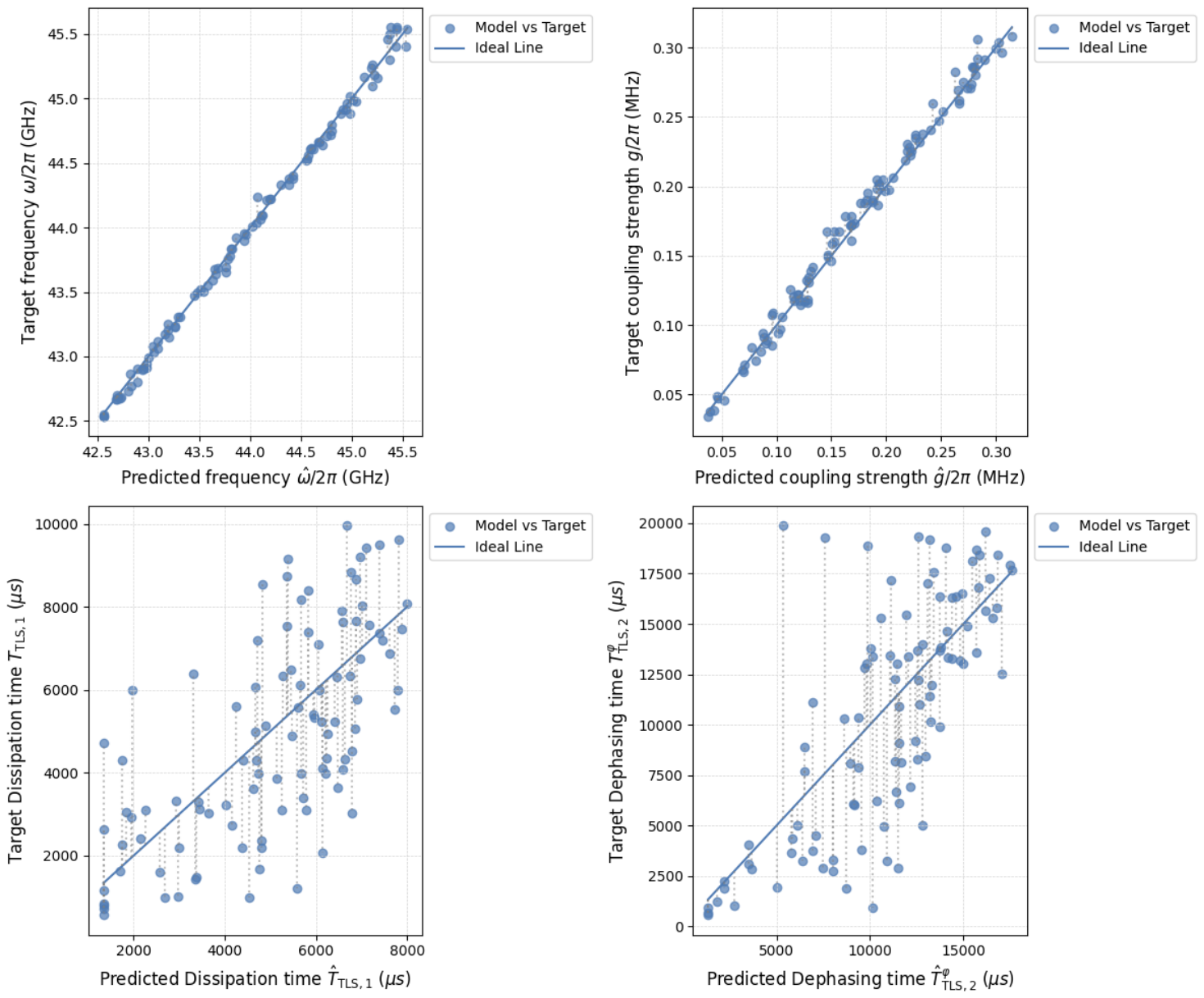} 
     \caption{\textbf{Model M1: Predictions vs. Targets.} Scatter plots comparing predicted values with ground-truth targets on 100 samples from the test set for a) the TLS frequencies $\omega_\text{TLS}$, (b) the coupling strengths $g$, (c) The dissipation time \(T_{\text{TLS}, 1}\) and (d) The dephasing time \(T^\varphi_{\text{TLS}, 2}\)}
     
     \label{fig:mp_without_noise}
\end{figure}

In the case of \textbf{M1} model, the model predictions are highly accurate for $\omega_{\text{TLS}}, g$. This can be confirmed from the Figure \ref{fig:mp_without_noise} which shows a near-perfect linear correlation between the predicted and expected values. In case of $T_{\text{TLS}, 1}$ and $T^\varphi_{\text{TLS}, 2}$, the predictions show a greater deviation from the ideal line indicating a reduction in performance. This observation is further supported by the MSE loss on the test set, as shown in table \ref{tab:mse_values}.

\begin{table}[htbp]
\centering

\caption{Mean squared error (MSE) loss values on the test dataset}

\label{tab:mse_values}
\begin{tabular}{|c|p{3.2cm}|p{3.2cm}|p{3.2cm}|}
\hline
\textbf{Parameter} & \textbf{Model(M1) MSE loss value} & \textbf{Model(M2) MSE loss value} & \textbf{Model(M3) MSE loss value}\\
\hline
$\omega$           & 0.003  & 8461.5 & 0.004 \\
$g$                &  0.00001  &   4.91 & 0.00003 \\
$T_{\text{TLS}, 1}$   & 4672253.5 & 3696050176.0 & 6588405.5 \\
$T^\varphi_{\text{TLS}, 2}$ & 16321636.0  & 257644380160.0 & 19207252.0 \\
\hline
\end{tabular}
\end{table}

\begin{figure}[h]
    \centering
    \includegraphics[width=0.8\linewidth]{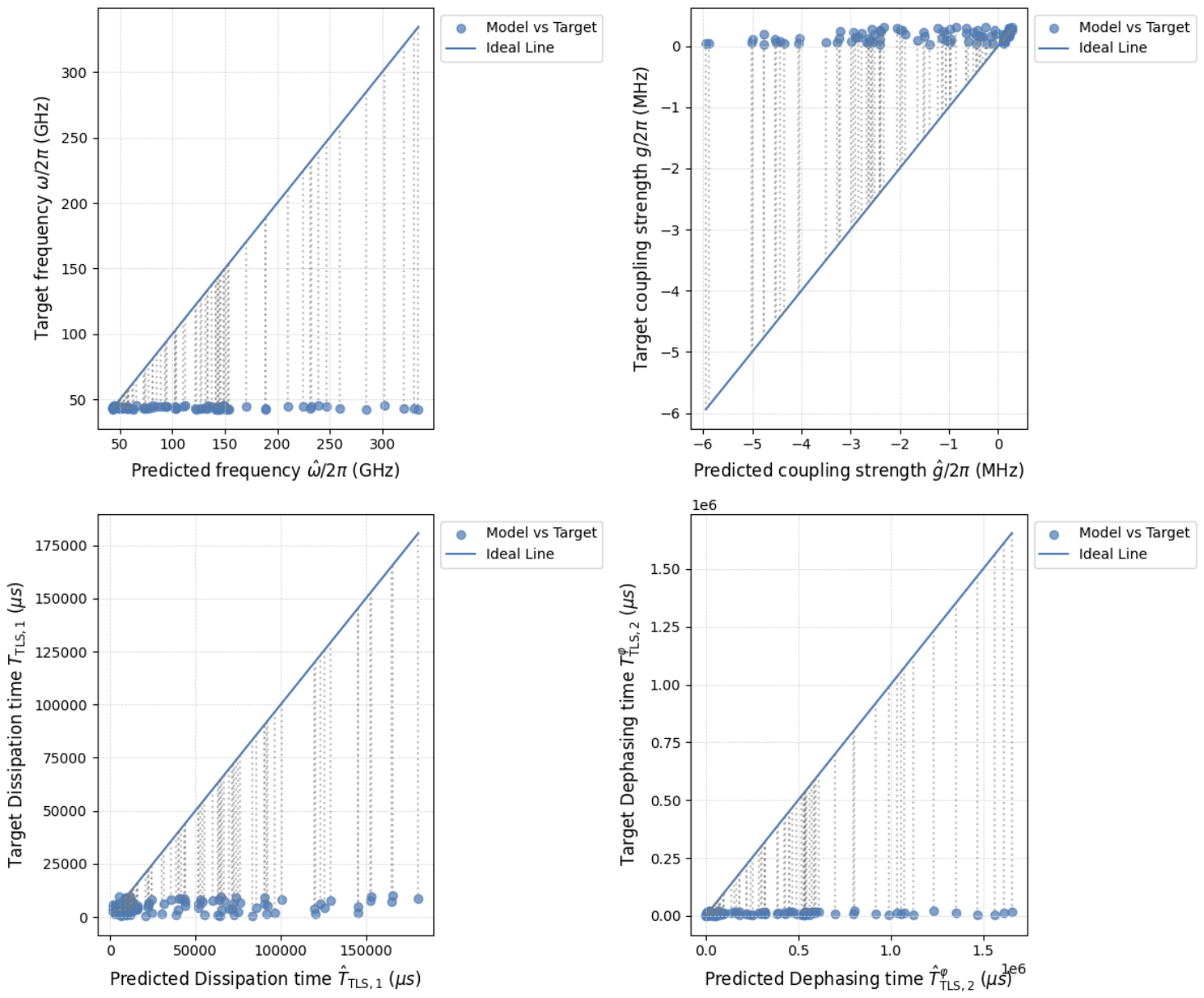} 
    \caption{\textbf{Model M2: Predictions vs. Targets.} Scatter plots comparing predicted values with ground-truth targets on 100 samples from the test set for a) the TLS frequencies $\omega_\text{TLS}$, (b) the coupling strengths $g$, (c) The dissipation time \(T_{\text{TLS}, 1}\) and (d) The dephasing time \(T^\varphi_{\text{TLS}, 2}\)}
     \label{fig:mp_with_noise}
\end{figure}

\begin{figure}[h]
    \centering
    \includegraphics[width=0.8\linewidth]{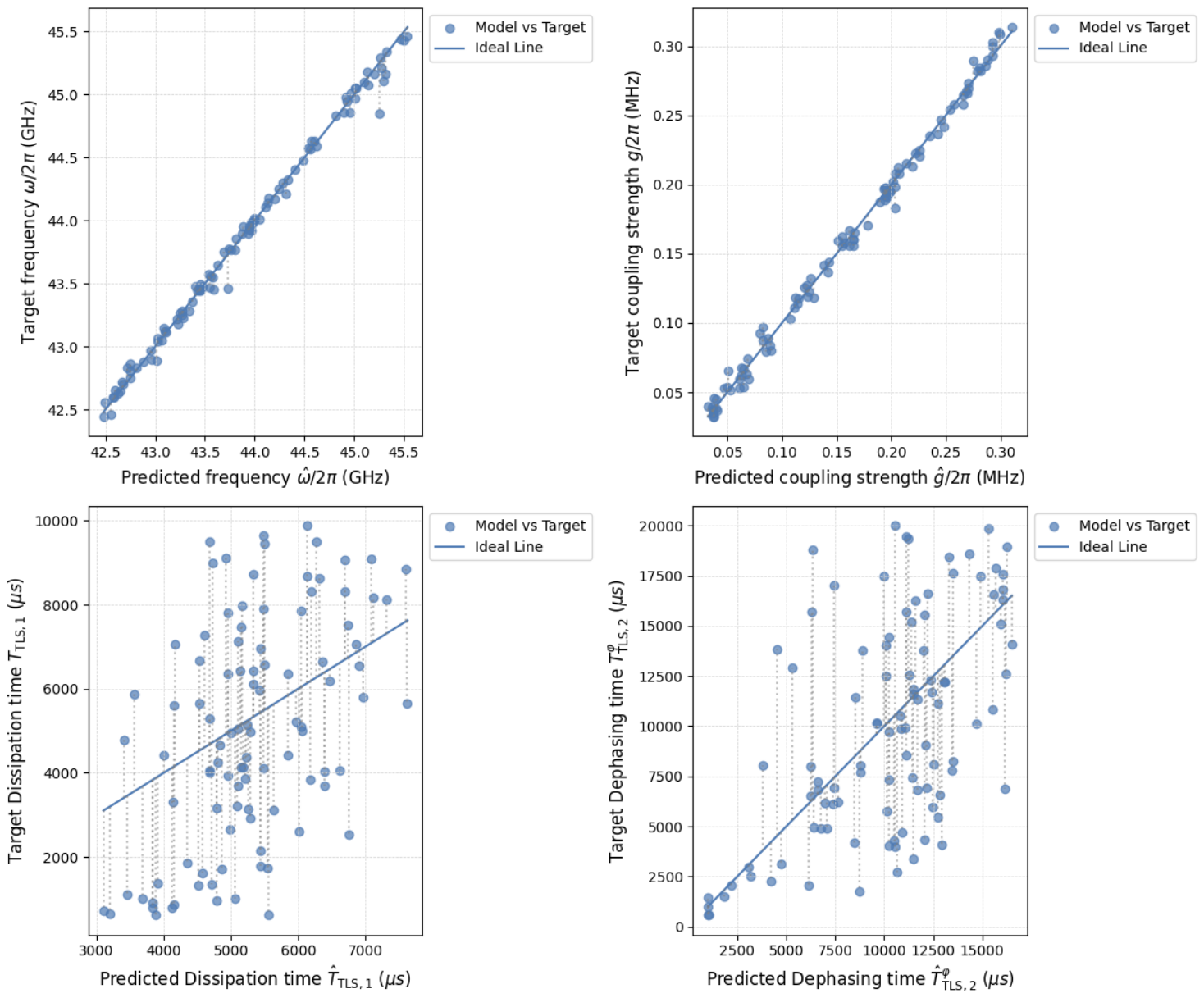} 
    \caption{\textbf{Model M3: Predictions vs. Targets.} Scatter plots comparing predicted values with ground-truth targets on 100 samples from the test set for a) the TLS frequencies $\omega_\text{TLS}$, (b) the coupling strengths $g$, (c) The dissipation time \(T_{\text{TLS}, 1}\) and (d) The dephasing time \(T^\varphi_{\text{TLS}, 2}\)}
     \label{fig:mp_trained_with_noise}
\end{figure}

For model  \textbf{M2}, there is a significant decline in performance across all four components of $\vec{q}$. This is evident from Figure \ref{fig:mp_with_noise} where the predictions $\vec(\hat{q})$ deviate significantly from the ideal line($\vec{q} = \vec\hat{q}$). To mitigate the impact of noise, the model \textbf{M3} was trained using a dataset that included representative samples of noisy image. This considerably improved the model performance, yielding results(shown in Figure \ref{fig:mp_trained_with_noise}) that are comparable to the \textbf{M1} model.

\subsection{Comparison to the analytic predictions}
To establish the usefulness of the CNN approach, we compare the results of the CNN model with those obtained via perturbation theory of the analytical expressions for the TLS frequency $\omega_\text{TLS} $and coupling strengths $g$. We use an automated procedure for estimating the parameters, which works as follows. We take the difference of the maximum and mean populations for each value of the drive frequency $\omega_d$, and pick the two highest values near the correct qubit and TLS frequencies $\omega_q$ and $\omega_\text{TLS}$ taken from the labels. From this, we get the shifted detuning $\tilde \Delta = \tilde \omega_\text{TLS} - \tilde \omega_q$. Next, we take the vertical slice for the drive frequency at the shifted TLS frequency $\tilde \omega_\text{TLS}$. Then, we fit the function 
\begin{equation}\label{eq:fit}
    p(t) = c_1 + c_2 e^{-c_3 t} \cos{c_4 t + c_5},
\end{equation}
where the fitting parameter $c_4 = \Omega$ corresponds to the Rabi frequency of the driven transition. With the approximate shifted detuning, Rabi frequency, and the known values of the drive amplitude $A$ and the anharmonicity $U$, we can estimate the coupling strength from
\begin{equation}
    g = \frac{\Omega \tilde \Delta}{A}
\end{equation}
and the TLS frequency from
\begin{equation}
    \omega_\text{TLS} = \tilde \omega_\text{TLS} + \frac{g^2}{\tilde \Delta}.
\end{equation}
Since the automated procedure does not perfectly pick the peaks corresponding to the correct TLS or qubit frequencies, or fit the function Eq.~\eqref{eq:fit}, we pick a set of 200 samples for which the procedure works based on a visual inspection. This is to avoid comparing the CNN results to estimates where the procedure fails. The mean square errors for the estimates on the 200 samples are $\text{MSE}_\omega = 0.000831$ and $\text{MSE}_g = 0.0138$, for the TLS frequency $\omega_\text{TLS}$ and the coupling strength $g$, respectively.

\begin{figure}
    \centering
    \includegraphics[width=\linewidth]{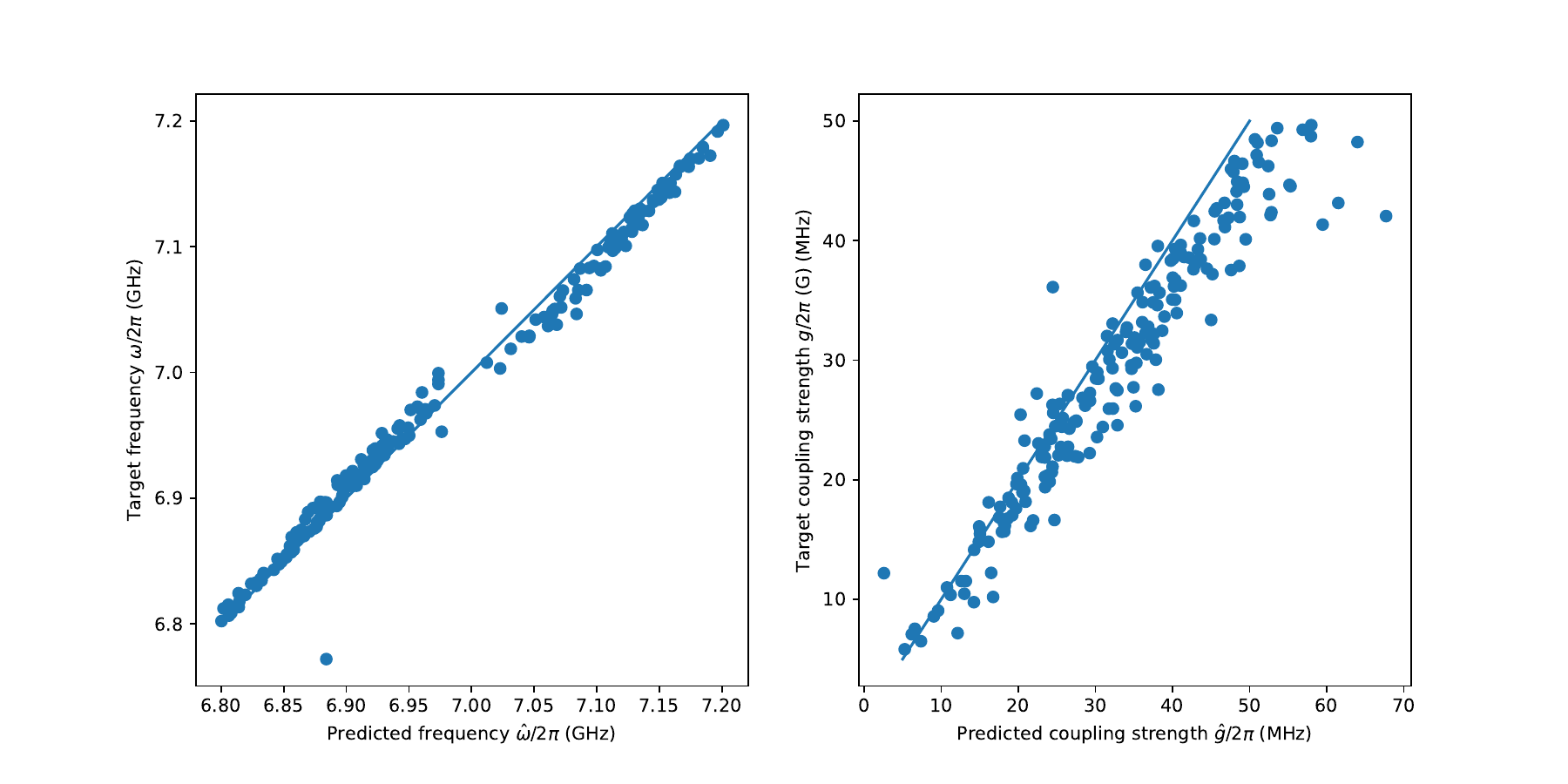}
    \caption{The analytical estimates vs Target (a) the TLS frequencies $\omega_\text{TLS}$ and (b) the coupling strengths $g$. }
    \label{fig:analytical_estimates}
\end{figure}

Seeing that the analytical expressions are based on perturbation theory, i.e., only valid when $g \ll \Delta$, we expect that the analytical expression give the largest errors when the detuning $\Delta$ is small. This can be seen in Fig.~\ref{fig:analytical_estimates} (a). Clearly, this method for estimating the parameters does not work for samples where the TLS frequency is outside the frequency range $\omega \in [\omega_q - \SI{0.2}{\giga\hertz}, \omega_q + \SI{0.2}{\giga\hertz}]$, a limitation not necessarily shared by the CNN model; we omit such samples from the set.

\section{Discussion}
We have developed a deep learning model that can extract the TLS parameters from spectroscopic data containing a single TLS more efficiently and rapidly than traditional perturbation theory methods. Notably this procedure is successful regardless of whether noise is introduced into the system. In case of noise, training the model on a dataset containing samples with noise significantly improves its performance, enabling it to achieve results comparable to those obtained in the clean limit. The natural application of this protocol would be fast-throughput spectroscopic analysis of qubits for characterization and quality assessment of multiple qubit devices. 

For future work, we propose exploring alternative model architectures and conducting extensive hyperparameter tuning to further enhance performance and robustness, as well as extending this approach to enable the extraction of qubit parameters in the presence of multiple strongly coupled TLS.

\section{Acknowledgements}
Work  of AB, AT and AP was supported by 
AFOSR program FA9550-25-1-0103, work of OM was supported by Novonordisk Quantum Computing Program. 
We are grateful to Josu Gomez Beldarrain, Alex Bilmes, Ilya Drozdov, Vito Iaea,  Cameron Kopas,  Kasper Grove-Rasmussen, Morten Kjaergaard, Peter Krogstrup, Vincent Michal,  Thue Christian Thann Nikoenlajs,  Dave Pappas, Yaniv Rosen and  Mark Svendsen  for useful discussions. We thank the Center for Information Technology of the University of Groningen for their support and for providing access to the Hábrók high performance computing cluster.

\bibliography{ref}
\end{document}